\documentclass{aa}
\usepackage[utf8]{inputenc}
\usepackage{graphicx}
\usepackage{array}
\usepackage{txfonts}
\usepackage{natbib}
\usepackage{booktabs,caption,fixltx2e}
\usepackage[flushleft]{threeparttable}
\bibpunct{(}{)}{;}{a}{}{,}

\title{Charge exchange in galaxy clusters}
\author{ Liyi Gu \inst{1,2} 
\and 
Junjie Mao \inst{2,3} 
\and 
Jelle de Plaa \inst{2}
\and
A.J.J. Raassen \inst{2,4}
\and
Chintan Shah \inst{5}
\and
Jelle S. Kaastra \inst{2,3}}

\offprints{Liyi Gu (l.gu@sron.nl)}
\date{\today}

\institute{RIKEN Nishina Center, 2-1 Hirosawa, Wako, Saitama 351-0198, Japan
\and
SRON Netherlands Institute for Space Research, Sorbonnelaan 2,
           3584 CA Utrecht, the Netherlands 
           \and 
Leiden Observatory, Leiden University, PO Box 9513, 2300 RA Leiden, the Netherlands 
\and 
Astronomical Institute ``Anton Pannekoek'', Science Park 904, 1098 XH Amsterdam, University of Amsterdam, The Netherlands
\and
Max-Planck-Institut f$\rm \ddot{u}$r Kernphysik, Heidelberg, D-69117 Heidelberg, Germany
    }

\abstract{Though theoretically expected, the charge exchange emission from galaxy clusters has never been confidently detected. Accumulating
hints were reported recently, including a rather marginal detection with the \textit{Hitomi} data of the Perseus cluster. As suggested in \citet{gu2015}, a detection of charge exchange line emission
from galaxy clusters would not only impact the interpretation of the newly-discovered 3.5 keV line, but also open up a new research topic on the interaction
between hot and cold matter in clusters. }
{
We aim to perform the most systematic search for the \ion{O}{VIII} charge exchange line in cluster spectra using the RGS on board \textit{XMM-Newton}. 
}{
We introduce a sample of 21 clusters observed with the RGS. In order to search for \ion{O}{VIII} charge exchange, the sample selection criterion is a $>5\sigma$ detection of the \ion{O}{VIII} Ly$\alpha$ line in the archival RGS spectra. The
dominating thermal plasma emission is modeled and subtracted with a two-temperature CIE component, and the residuals are stacked for the line search. 
The systematic uncertainties in the fits are quantified by refitting the spectra with a varying continuum and line broadening.
}{
By the residual stacking, we do find a hint of a line-like feature at 14.82~{\AA}, the characteristic wavelength expected for oxygen charge exchange.
This feature has a marginal significance of $2.8\sigma$, and the average equivalent width is $2.5 \times 10^{-4}$ keV. We further demonstrate that the putative feature can be hardly affected by the systematic errors from continuum modelling and instrumental effects, or the atomic uncertainties of the neighbouring thermal lines.
}{
Assuming a realistic temperature and abundance pattern, the physical model implied by the possible oxygen line agrees well with the theoretical model proposed previously to explain the reported 3.5~keV line. If the charge exchange source indeed exists, we would expect that the oxygen abundance is potentially overestimated by $8-22$\% in previous X-ray measurements which assumed pure thermal lines. This new RGS results bring us one step forward to understand the charge exchange phenomenon in galaxy clusters.
}
\titlerunning{Charge exchange in galaxy clusters}
\authorrunning{L. Gu}

\begin{document}

\maketitle

\section{Introduction}

Charge exchange (hereafter CX) occurs when a neutral atom collides with a sufficiently charged ion, and recombines the ion into
a highly-excited state. It is the dominant atomic process at the interface where the highly-charged solar wind interacts with 
comet atmospheres \citep{lisse1996, cravens1997}. X-ray observations showed that the solar wind CX also influences planet atmospheres and the heliosphere \citep{snowden2004, dennerl2006, fujimoto2007, br2007, smith2014}. Although the observation of CX from extrasolar objects is still quite challenging, there are a growing number of reports for possible CX from supernova remnants \citep{katsuda2011} starburst galaxies \citep{tsuru2007, liu2011}, active galactic nuclei \citep{gu2017}, and galaxy clusters \citep{fabian2011, gu2015, hitomi3_5}. 

Among these objects, the possible CX from galaxy clusters recently attracts more attention, because it becomes intimately related with the interpretation of a newly-discovered potential X-ray line at $\sim 3.5$ keV. As reported in \citet{bulbul2014},  \citet{boyarsky2014}, and more recently in \citet{cappe2017}, a weak line feature at $\sim 3.5$ keV is detected in the spectra of a large sample of galaxies and clusters observed by \textit{XMM-Newton} EPIC and \textit{Chandra} ACIS. It cannot be identified immediately in the standard line tables of thermal plasma. These authors thus considered it as a truly exceptional feature, and proposed that it might stem from particle physics $-$ radiative decay of a dark matter candidate called sterile neutrino. However, in \citet{gu2015}, we showed that there might be a missing element in their atomic model: the charge exchange in the hot intracluster medium (hereafter ICM). The observed residual at $\sim 3.5$ keV with the X-ray CCDs would be washed away (or reduced to within 1$\sigma$) by introducing the CX-excited \ion{S}{XVI} line at $\sim 3.45$ keV. Our atomic calculation has been verified recently by a dedicated laboratory measurement \citep{shah2016}. 

Not only the physical interpretation, but also the detection of the 3.5 keV line itself is still in controversy. So far all the detections were made by X-ray CCDs with spectral resolution of $\sim 100$~eV. The putative line is then blurred into a 1\% bump above the continuum, affected easily by instrumental and atomic uncertainties. Resolving the feature with a high resolution spectrometers is therefore essential. The calorimeter on board the \textit{Hitomi} satellite provides a resolution of $\sim 5$~eV at the target energy, offering for the first time such an opportunity. As reported in \citet{hitomi3_5}, the \textit{Hitomi} Perseus data can rule out, at least at $99$\%, an unidentified line at the flux level reported in the previous \textit{XMM-Newton} and \textit{Chandra} studies, even though the \textit{Hitomi} data were a bit shallow in the target energy band. Interestingly, it does show a hint for the \ion{S}{XVI} CX line at $\sim 3.45$ keV. By analyzing the \textit{Hitomi} data in more detail, a hint was discovered of the \ion{Fe}{XXV} CX line at $\sim 8.78$ keV \citep{hitomi-atomic}, and the upper limits of the sulfur and iron line fluxes are well in line with the model prediction in \citet{gu2015}.

The \textit{Hitomi} results open up a new window for CX astrophysics. The possible CX signal from galaxy clusters provides a new approach to locate the cold matter and to investigate its interaction with the hot ICM. To confirm the \textit{Hitomi} results,
we carry out a systematic CX line search using the existing cluster data obtained with the \textit{XMM-Newton} Reflection Grating Spectrometer (RGS), which offers a high resolving power of $150-700$ for soft X-ray lines from galaxy clusters. We aim to achieve the most stringent
constraint on the CX phenomenon in clusters using the current state-of-the-art of high-resolution X-ray spectrometers. This paper is structured as follows: first we determine the target CX line by a theoretical calculation in Sect.~\ref{sect:target}; in Sect.~\ref{sect:sample} and Sect.~\ref{sect:data}, we present the target selection and data reduction. Sect.~\ref{sect:analysis} describes the analysis and shows the results based on the RGS sample. The physical implications of the observed results are presented in Sect.~\ref{sect:discussion} and summarized in Sect.~\ref{sect:conclusion}. Throughout the paper, the errors are given at a 68\% confidence level.

\begin{table*}[!htbp]
\centering
\caption{\textit{XMM-Newton} RGS data of the sample clusters}
\label{tab:sample}
\begin{threeparttable}
\begin{tabular}{lcccccccccccccccc}
\hline
cluster & Observation ID & Total clean time (ks) & $kT^{(a)}$ (keV) & $z$ & O abundance  \\
\hline 
2A 0335+096 & 0109870101 0147800201              & 91.6  & 4.0  & 0.0349   & 0.59 \\
A85 & 0723802101/2201                            & 155.6 & 6.1  & 0.0557   & 0.55   \\
A133 & 0144310101 0723801301/2001                & 136.0 & 3.8  & 0.0569   & 0.67   \\
A262 & 0109980101 0504780201                     & 80.3  & 2.2  & 0.0161   & 0.56   \\
A383 & 0084230501                                & 22.3  & 3.1  & 0.1871   & 0.38   \\
A496 & 0095010901 0135120201/0801 0506260301/0401 & 152.4& 4.1  & 0.0328   & 0.60   \\
A1795 & 0097820101                               & 31.8  & 6.0  & 0.0616   & 0.35   \\
A1991 & 0145020101                               & 37.8  & 2.7  & 0.0586   & 0.65   \\
A2029 & 0111270201 0551780201/0301/0401/0501     & 156.7 & 8.7  & 0.0767   & 0.41   \\
A2052 & 0109920101/0201/0301 0401520301/0501/0601 & 153.5 & 3.0  & 0.0348   & 0.52   \\
      & 0401520801/0901/1101/1201/1301/1601/1701  &      &      &          &        \\
A2199 & 0008030201/0301/0601 0723801101/1201     & 118.6  & 4.1  & 0.0302   & 0.62   \\
A2204 & 0112230301 0306490101/0201/0301/0401     & 77.0  & 5.6  & 0.1511   & 0.28   \\
A2597 & 0147330101 0723801601/1701               & 175.8 & 3.6  & 0.0852   & 0.54   \\
A3112 & 0105660101 0603050101/0201               & 162.2 & 4.7  & 0.0750   & 0.51   \\
A3581 & 0205990101 0504780301/0401               & 119.8 & 1.8  & 0.0214   & 0.47   \\
A4038 & 0204460101 0723800801                    & 67.1  & 3.2  & 0.0283   & 0.66   \\
A4059 & 0109950101/0201 0723800901/1001          & 200.7 & 4.1  & 0.0460   & 0.58   \\
AS 1101 & 0123900101 0147800101                  & 124.2 & 3.0  & 0.0580   & 0.32   \\
EXO 0422 & 0300210401                            & 29.6  & 3.0  & 0.0390   & 0.65   \\
Hydra-A & 0109980301/0501 0504260101             & 118.0  & 3.8  & 0.0538   & 0.35   \\
ZW 3146 & 0108670101 0605540201/0301             & 166.2 & 3.6  & 0.2906   & 0.45   \\

\hline
\end{tabular}
\begin{tablenotes}
\item[$(a)$] Temperatures, redshifts, and oxygen abundances are taken from \citet{deplaa2017},
except for Abell~383, Abell~2204, and ZW~3146. The best-fit values obtained with single-temperature
modelling (Sect.\ref{sect:analysis}) are reported for these three objects.
\end{tablenotes}
\end{threeparttable}
\end{table*}

\section{Target charge exchange line \label{sect:target}}

\begin{figure}[!htbp]
\resizebox{\hsize}{!}{\includegraphics[angle=0]{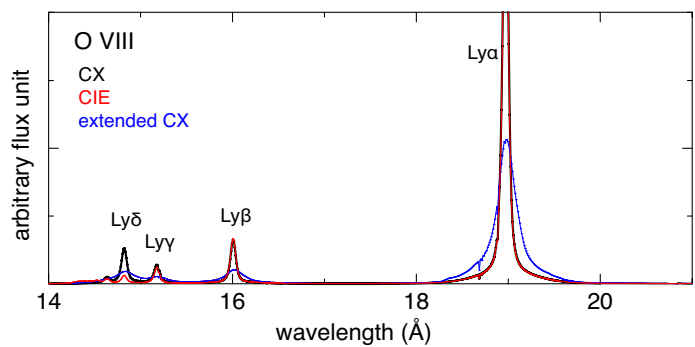}}
\caption{CX (black) and CIE (red) \ion{O}{VIII} lines convolved with the RGS response. The plasma temperature is set to 2~keV, and 
the collision velocity for CX is 500 km~s$^{-1}$. The two models are normalized
at the Ly$\alpha$ line. The blue curve shows the CX lines further broadened by the spatial extent of a representative cluster (2A~0335). }
\label{fig:cx-cie}
\end{figure}

This work aims to search for CX signatures using the RGS instrument, which already limits the candidate elements to be C, N, O, Ne, Mg, Fe, and Ni.
For the ICM, C and N are hard to detect, and the wavelengths of the \ion{Mg}{XII} CX lines ($\sim$6.4~{\AA}) are too close to
the bandpass limit. For the Fe and Ni L-shell lines (mostly Li-like to Na-like sequences), the theoretical calculations for CX cross sections
are still lacking, albeit with a few laboratory measurements \citep{beiersdorfer2008}. The best candidate thus becomes O, which is generally much more abundant than Ne. We would focus on \ion{O}{VIII} recombined 
from proton-like oxygen ions, which are the most abundant species of oxygen ions in the ICM.

To determine the target emission feature, we calculate the CX lines using the \textit{cx} model in SPEX version 3.03.00 \citep{gu2016},
and compare them with the CIE emission (Fig.~\ref{fig:cx-cie}). The model plasma has a temperature of 2~keV and solar abundance. 
We assume CX collisions between bare oxygen and hydrogen atoms at a velocity of 500 km~s$^{-1}$. When normalizing the CX and CIE lines
at the Ly$\alpha$ transitions, the only significant difference appears at the Ly$\delta$ (1s$-$5p) lines at $\sim 14.82$~{\AA}. This is because 
electron capture into atomic levels with principal quantum number $n=5$ is dominant for the adopted collision condition. In fact, according 
to the theoretical calculation by e.g., \citet{janev1993}, $n=5$ remains as the peak capture channel for a collision velocity $\leq$ 4000 km~s$^{-1}$, 
which means that the \ion{O}{VIII} Ly$\delta$ transitions would be the key CX signature in most of the relevant astrophysical conditions. 
A more recent $n-$resolved multi-channel Landau-Zener calculation by \citet{mullen2017} shows the same peak at $n = 5$ for a similar velocity range.

\section{Sample selection \label{sect:sample}}

The cluster sample is selected using three criteria. First, the targets must show significant \ion{O}{VIII} Ly$\alpha$ lines.
A quick filtering is done by scanning through the standard RGS spectra archived in the BiRD catalog\footnote{http://xmm.esac.esa.int/BiRD/},
and picking up spectra with \ion{O}{VIII} Ly$\alpha$ lines that can be identified by eye. Then we extract the RGS 
spectra of candidates (see the data reduction in Sect.~\ref{sect:analysis}), and determine the \ion{O}{VIII} Ly$\alpha$ line significance
by fitting the restframe $0.62-0.69$ keV band with two Gaussians representing Ly$\alpha$1 and Ly$\alpha$2, and the line-free CIE model
for the continuum. Then we selected objects with \ion{O}{VIII} Ly$\alpha$ line significance higher than 5$\sigma$. This reduces the catalog 
to a list of 49 objects, which forms the ``preliminary sample''. 

Second, the nearby M87 and Perseus cluster are removed from the sample for their oversized angular extents, which blur the target line
and reduce the detection sensitivity. The remaining number is 47, to be called ``intermediate sample''.

The third criterion is to exclude objects with strong \ion{Fe}{XVII} emission lines. The \ion{Fe}{XVII} line at 15.02~{\AA} is a close neighbour
to the target \ion{O}{VIII} Ly$\delta$ line at 14.82~{\AA}, so that the astrophysical and instrumental broadening of the Fe line might potentially affect the 
detection of the O line. Using the updated ionization equilibrium balance calculation in SPEX 3.03.00 \citep{u17}, we find that the absolute concentration of \ion{Fe}{XVII} 
becomes negligible ($< 10^{-10}$) at a balance temperature of 1.8 keV, which is then set as the final criterion for the sample. We check the 
average temperatures reported in \citet{pinto2015} and keep objects with temperatures higher than 1.8~keV. This reduces the ``intermediate sample'' to 
the final sample of 21 objects. The properties of the 21 objects are listed in Table~\ref{tab:sample}.

\section{Data \label{sect:data}}

We process the XMM-Newton RGS and MOS data, mostly following the method described in \citet{pinto2015}. The MOS data are not used for spectral analysis, but for determining the spatial extent of the source along the dispersion direction of the RGS.

The data reduction is done with the SAS version 16.0.0 and the latest calibration files. The SAS tasks \textit{rgsproc} and \textit{emproc} are run for
the RGS and MOS data, respectively. The time intervals contaminated by soft-proton are detected using the light curves of the RGS CCD9,
which are then filtered by a 2$\sigma$ clipping. To search extensively for the CX signal, the
width of the RGS source extraction region (\textit{xpsfincl}) is set to be 99\% of the point spread function, which is approximately 
a 3.4-arcmin-wide belt centered on the emission peak. The modelled background spectra are used in the spectral analysis. The diffuse cosmic background
is not modelled explicitly, since it is smeared out and merged into the continuum. The spectral files are 
converted to SPEX format through the SPEX task \textit{trafo}. 

To model the spectral broadening due to the spatial extent of the sources, we extract the MOS1 image in the $0.5-1.8$~keV band for each observation,
and calculate the surface brightness profile in the RGS dispersion direction using the \textit{rgsvprof} task in SPEX. The spectral broadening is then modelled using the SPEX model \textit{lpro}
based on the brightness profiles. The line broadening can be further fine-tuned by varying the scale parameter \textit{s} of the \textit{lpro} 
model, which is left free in the fitting.

We analyze the first and second order RGS spectra in the $8-28$~{\AA} band. All abundances are relative to \citet{lodders2009} proto-solar standard. We
use optimally binned spectra by the \textit{obin} command in SPEX \citep{kaastra2016}, and C-statistics for fitting and error estimation. We adopt the updated 
ionization balance calculations of \citet{u17}.

\section{Spectral analysis and results \label{sect:analysis}}

We aim to search for a weak line feature at 14.82~{\AA}, which means that the dominant thermal emission must be properly modelled and subtracted. 
Here we carry out two approaches, the global and local fits, to model the thermal emission. The global fit uses a full self-consistent calculation of 
line and continuum emission to fit the entire band, and the local fit calculates the sum of a simple continuum plus a series of Gaussian emission lines
to model a narrow band near the target line. The latter method might provide a better representation of the local continuum, while it cannot model all the 
weak lines, especially the thermal \ion{O}{VIII} Ly$\delta$ component at the target wavelength. Thus the two approaches would compensate each other. Combined global and local fits were also used in other recent X-ray work, e.g., \citet{mernier2016}.

\subsection{Global fits}

\begin{figure*}[!htbp]
\resizebox{0.6\hsize}{!}{\includegraphics[angle=0]{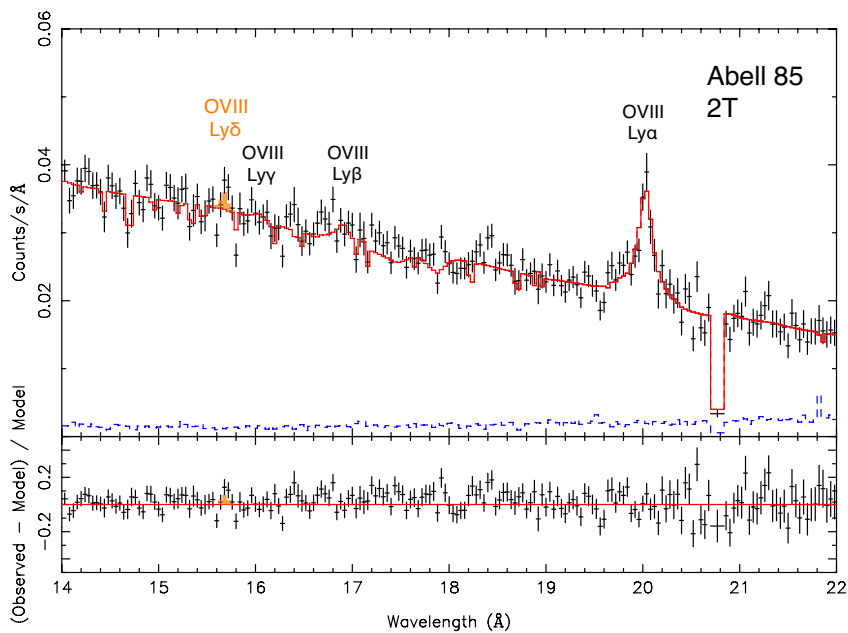}}
\caption{Example RGS 1st order spectrum of Abell~85. The best-fit two-temperature model is shown in red, and the subtracted instrumental
background is plotted in blue. The target \ion{O}{VIII} Ly$\delta$ feature is marked by orange.}
\label{fig:a85}
\end{figure*}

To describe the cluster thermal component, we fit the spectra using two SPEX \textit{cie} models. This is because our sample is built-up from mostly cool-core clusters, which are known to show both hot- and cool- phase ICM in the cores \citep{gu2012}. The emission measures and electron temperatures of the hot and cool components are left to vary, while the N, O, Ne, Mg, Fe, and Ni abundances of the two \textit{cie} components are coupled to each other. The ion temperatures are tied to the electron temperatures. The two-temperature model is modified first by a \textit{redshift} component,
then by a \textit{hot} model for foreground absorption. The absorber is neutral (kT = 0.5~eV), has solar abundances, and its column density is set to the calculated value from \citet{willingale2013}, which includes the contribution from both atomic and molecular components. Finally, the two-temperature model is multiplied by the \textit{lpro} component to correct for the spatial broadening of the objects (Sect.\ref{sect:data}). All the spectra in our sample can be modelled with pure thermal emission, none of them requires an additional power-law spectral component for the AGN. The same treatment was also used in \citet{pinto2015} and \citet{deplaa2017}. This does not necessarily mean that AGNs are absent, but merely that the thermal components are apparently dominant in these spectra. The global fit of the RGS spectrum of Abell~85 is shown in Fig.~\ref{fig:a85} as an example.

To measure the intensity of the possible weak line at 14.82~{\AA}, we further add a Gaussian component with a fixed wavelength in the fits. The Gaussian line is multiplied by the same redshift 
and absorption components as those applied to the thermal model. The spatial distribution of the possible charge exchange component is unclear, it might be either emitted mostly from the cluster
core, or significantly contributed by the cool gas in member galaxies. In the former case, a point source PSF would be adequate for the line, while in the latter case, the Gaussian should be    
multiplied by the \textit{lpro} component of the entire cluster (Fig.~\ref{fig:cx-cie}). In Fig.~\ref{fig:ew}, we show the best-fit results for both cases.

\subsection{Local fits}

\begin{table}[!htbp]
\caption{Emission lines included in the local fits}
\label{tab:loclines}
\centerline{
\begin{tabular}{ll}
\hline\hline
Ion & Rest wavelength \\
    & ({\AA})  \\
\hline
\ion{O}{VIII} & 18.97  \ 16.00  \ 15.18 \\
\ion{Fe}{XVII} & 15.02 \ 15.26 \ 16.78 \ 17.05 \ 17.10 \\
\ion{Fe}{XVIII} & 14.20 \ 14.21 \ 14.37 \ 14.53 \ 15.63 \\
                & 15.83 \ 16.00 \ 16.07 \ 16.17 \ 17.61 \\
\ion{Fe}{XIX}   & 13.49 \ 13.52 \ 13.79 \ 14.66 \ 15.07 \\
                & 15.17 \ 16.06 \ 16.22 \\
\ion{Fe}{XX}    & 14.28 \ 14.77 \ 14.92 \\
\ion{Ne}{IX}    & 13.44 \ 13.55 \ 13.69 \\
\hline
\end{tabular}
}
\end{table}

To search for the target weak line, it is crucial to model the continuum and related thermal line to high precision. Although the global fits can describe the target band reasonably well, 
it might still be affected by the calibration imperfections of the RGS instrument \citep{cor2015}, the uncertainties in atomic code \citep{mernier2017}, and astrophysical complexity \citep{deplaa2017}. These
effects can be corrected by fitting the spectra locally, using a model including sufficient freedom to account for the thermal emission and the associated errors. The local fit is based on
a model including a single-temperature \textit{cie} continuum by setting \textit{ions} \textit{ignore} \textit{all} in SPEX, together with a set of Gaussian components with fixed central energies 
for the strong thermal lines.
The local fit is carried out in the 13~{\AA} $-$ 23~{\AA} band, and the added thermal emission lines are listed in Table~\ref{tab:loclines}. The continuum and Gaussian lines are  multiplied first by the \textit{redshift} and \textit{hot} components, and then by the \textit{lpro} for the spatial broadening. Another narrow Gaussian component is added at the cluster-frame wavelength of 14.82~{\AA} to account for the CX line. This approach is essentially the same as the one used in recent
weak line detection works (e.g., \citealt{bulbul2014}). 

The average C-statistics with the local fit is improved from the global fit by about 10 for $\sim 300$ degrees of freedom. Although useful to obtain a better
approximation to the 13~{\AA} $-$ 23~{\AA} band, the local method misses details from the global modelling of the full spectrum, for instance, the weak emission lines,
such as the thermal \ion{O}{VIII} Ly$\delta$ line, are ignored. The equivalent widths of the CX line obtained with the local fits must thus be treated as the upper limit.

\subsection{Results \label{sect:results}}

\begin{figure*}[!htbp]
\resizebox{0.6\hsize}{!}{\includegraphics[angle=0]{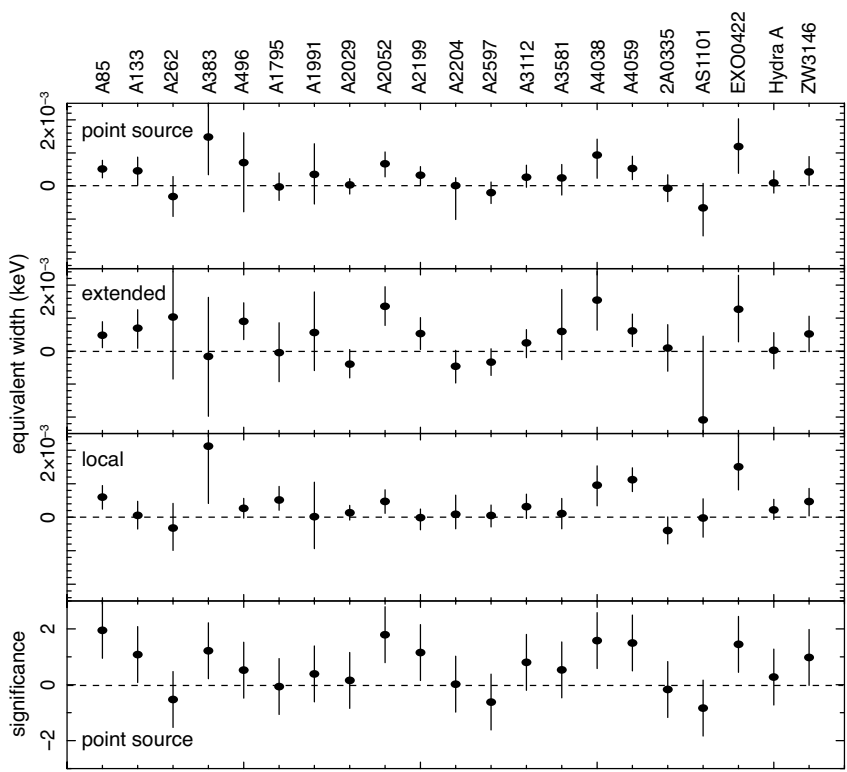}}
\caption{Best-fit equivalent widths of a Gaussian component at rest-frame wavelength of 14.82~{\AA}, based on both the 
global and local fits of the thermal component. For the global fits, the Gaussian line PSF is set to both the
point source and extended source modes. By dividing the best-fit equivalent width (global fits/point source) by their
errors, the line significance for each cluster is shown in the bottom panel. }
\label{fig:ew}
\end{figure*}

We calculate the equivalent widths of the Gaussian component at rest-frame 14.82~{\AA}, based on the best-fits from both global and local approaches. For the 
global fit, we further derive the equivalent widths for both point- and extended- source approximations. If the target line does not exist, the sample should show
positive and negative equivalent widths to be equally distributed around zero. As shown in Fig.~\ref{fig:ew}, the best-fit equivalent widths are apparently 
more seen on the positive side. The results using three methods (global/point source, global/extended source, and local/point source) agree well with each other, and
none of the sources has significantly negative line flux at 14.82~{\AA}. By dividing the best-fit equivalent widths by their errors, we further plot the significance
in Fig.~\ref{fig:ew}. It shows that all the seemingly excesses at 14.82~{\AA} are $\leq 2\sigma$ significance; most objects are consistent with 1$\sigma$.
This means that, although it does show a hint of a line at the target wavelength, it is not possible to report a significant detection in any individual object.

\begin{table*}[!htbp]
\caption{Weighted means of the best-fit equivalent widths}
\centering
\label{tab:eq}
\begin{tabular}{lcc}
\hline
Case & sample average  & uncertainties \\
& ($10^{-4}$ keV) & ($10^{-4}$ keV) \\
\hline 
Global, point source     & 2.5  & 0.9 \\
Global, extended         & 2.9  & 1.0 \\
Local, point source      & 2.8  & 0.9 \\
\hline
\end{tabular}
\end{table*}

\begin{figure*}[!htbp]
\resizebox{\hsize}{!}{\includegraphics[angle=0]{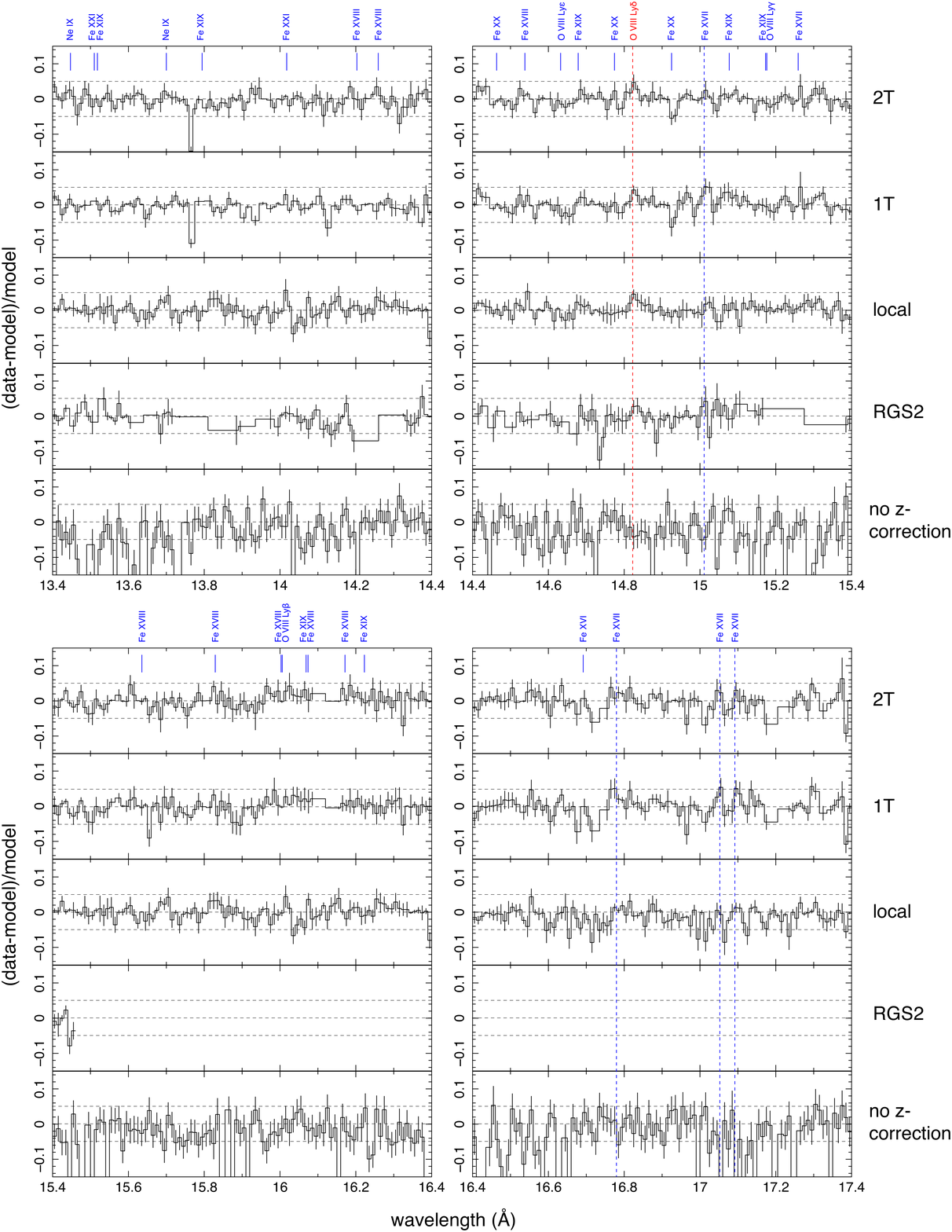}}
\caption{The stacked residuals in the rest-frame 13.4~{\AA} $-$ 17.4~{\AA} band. Different panels show the residuals from the two-temperature/sing-temperature global fits and the local fits, the residuals by RGS2 data alone, and the residuals
stacked without redshift correction. The dashed lines show a 5\% level. }
\label{fig:residual}
\end{figure*}

\begin{figure*}[!htbp]
\resizebox{\hsize}{!}{\includegraphics[angle=0]{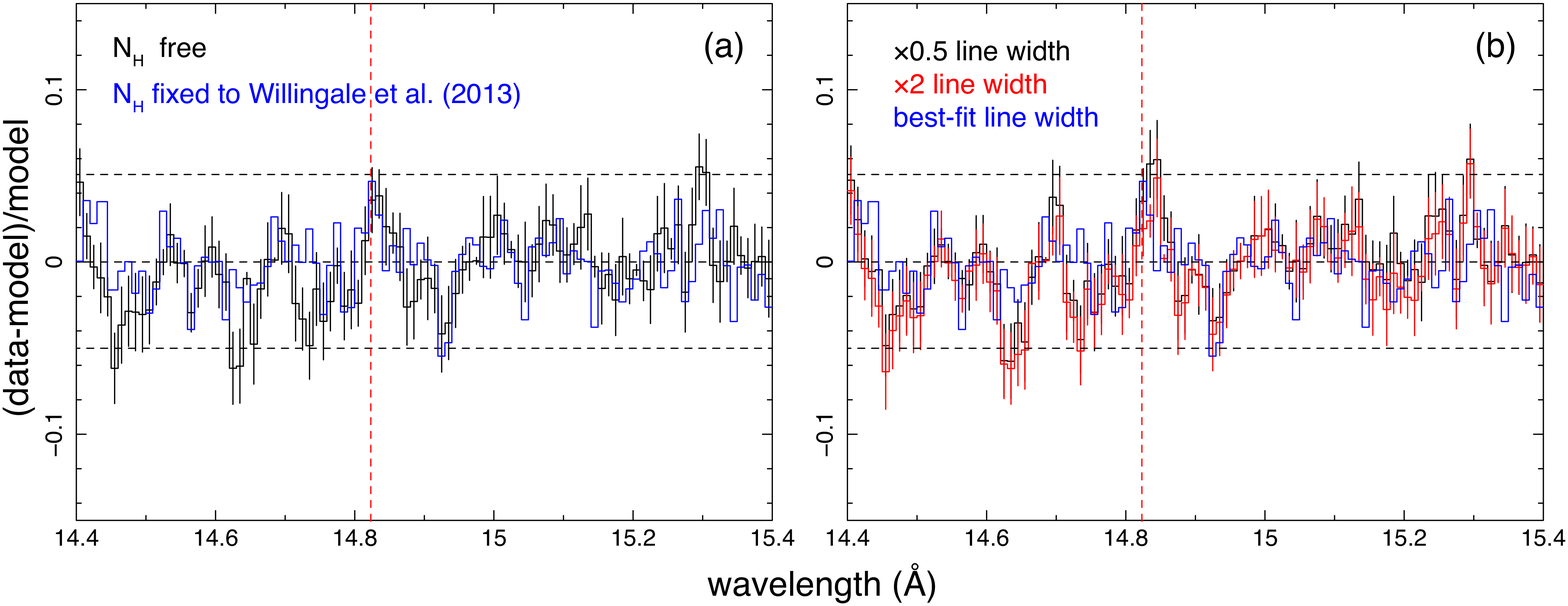}}
\caption{(a) Stacked residuals in the rest-frame 14.4~{\AA} $-$ 15.4~{\AA} band, yielded by fits with $N_{\rm H}$ free (black)
and with $N_{\rm H}$ values of \citet{willingale2013} (blue). (b) Results from the global fits with line broadening fixed to half (black) and two times (red) the best-fit value. The blue one is the same as shown in Fig.~\ref{fig:residual}. }
\label{fig:inst}
\end{figure*}

This naturally leads us to a stacking approach for better statistics. Here we blue-shift all the residual spectra to the source frame, and average them with a weighting based on 
the counts in each energy bin. The residual spectra for each object are obtained using the two-temperature, single-temperature, and local fit modellings. The Gaussian
component at 14.82~{\AA} is excluded for the stacking. In Fig.~\ref{fig:residual}, we plot the stacked residual spectra in the 13.4~{\AA} to 17.4~{\AA} band. The total stacking 
exposure is about 2.4~Ms. To examine instrumental
effects, we also show the residual plot based on the RGS2 data alone, as well as the one without redshift correction. The latter one is apparently dominated by instrumental
artifacts; by comparing it with the first three residuals in Fig.~\ref{fig:residual}, it is clear that the instrumental effects are smeared out by the redshift correction, while the features with cluster origin should stay. 

The stacked residuals of the two-temperature, single-temperature, and local fit modellings are in general agreement with each other. The amplitudes of the residuals are
mostly within 5\% of the model value; at 14.82~{\AA}, all the three stacked
plots consistently show a line-like excess, with a peak value of 3\%$-$4\%. By fitting the residual with a Gaussian line in the 14.8~{\AA} $-$ 14.85~{\AA} band, the significance of the target feature
is obtained to be 3.1$\sigma$, 2.8$\sigma$, and 3.4$\sigma$ with the two-temperature, single-temperature, and local modellings, respectively. The Gaussian line has a best-fit sigma of $0.03-0.04$~{\AA}, which agrees
with a narrow line feature broadened purely by the instrument (RGS FWHM $= 0.06-0.07$~{\AA} at the target wavelength). 
This feature can be seen using the RGS2 data alone, and the intensity is consistent with the RGS1+2 results, suggesting that it is unlikely to be an artifact on one of the two detectors. This is further supported by the residual spectrum stacked before the
redshift correction, in which no apparent instrumental feature (such as gaps) near the target wavelength is seen. 

Several other features are further revealed in Fig.~\ref{fig:residual}. The thermal \ion{Fe}{XVII} lines 
at rest-frame 15.02~{\AA}, 15.26~{\AA}, 16.78~{\AA}, 17.05~{\AA}, and 17.10~{\AA} are seen in the stacked residual based on single-temperature modelling; all the four lines are found at the correct wavelengths, indicating that the wavelength of the target feature at 14.82~{\AA} should also be accurate. 
The weak \ion{Fe}{XVII} features are at least partially modelled out by the two-temperature fit. This is because
the \ion{Fe}{XVII} lines must be emitted from the cool phase ICM (kT $<$ 1.5 keV), which is accounted for by the two-temperature model.

To be conservative, the stacked significance of the target line is 2.8$\sigma$ (single-temperature fit), which means a marginal detection. Since the wavelength
of the target line is well defined by atomic theory, and is accurately measured by experiments \citep{beiersdorfer2003}, there is no look-elsewhere effect \citep{gross2010} in the detection significance. 
As shown in Table~\ref{tab:eq}, the weighted-average of the equivalent widths derived from the global fits is 2.5 $\times 10^{-4}$ keV, with an upper limit of 3.4 $\times 10^{-4}$ keV. The detection of the 
putative feature might still be affected by several systematic uncertainties and biases, which will be studied as follows.

\subsection{Systematic effects}

 \subsubsection{Bias due to instrumental and astrophysical modelling}

\begin{figure*}[!htbp]
\resizebox{\hsize}{!}{\includegraphics[angle=0]{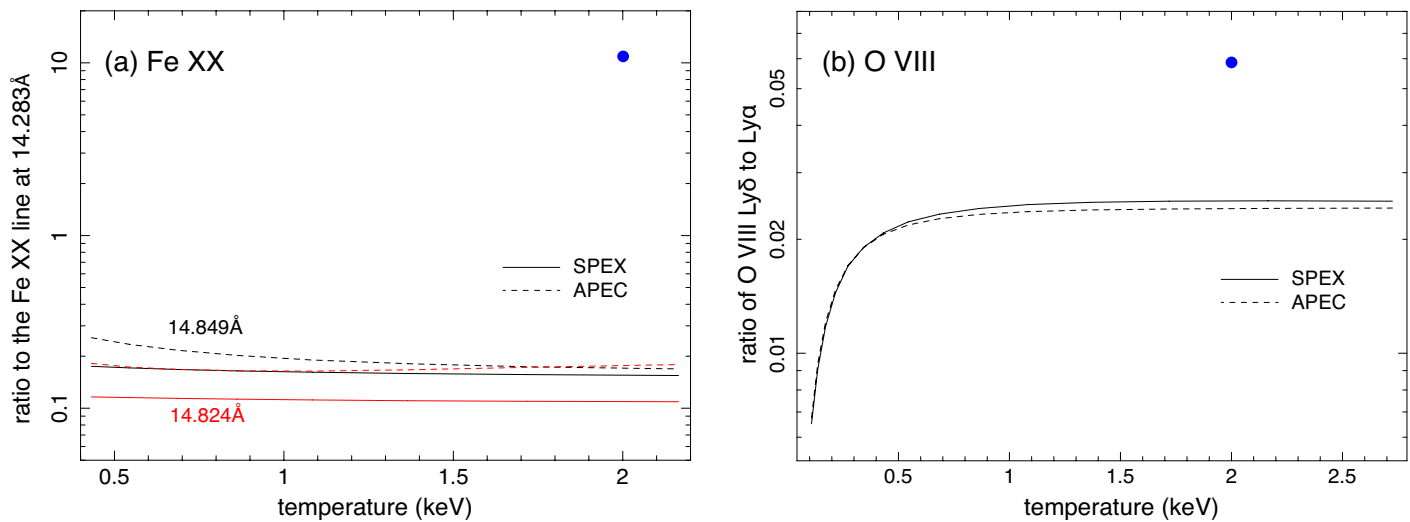}}
\caption{(a) Model line ratios of \ion{Fe}{XX} 14.849~{\AA} (black) and 14.824~{\AA} (red) to the main transition at 14.283~{\AA}, calculated
by the thermal model in SPEX v3.03 (solid) and APEC v3.0.8 (dashed). (b) \ion{O}{VIII} Ly$\delta$ to Ly$\alpha$ ratio as a function of temperature for the two thermal codes. The blue points in both figures show the value required to create a 1\% excess above the continuum at 14.82~{\AA} for a 2~keV plasma. }
\label{fig:fexx}
\end{figure*}

 The background component and AGN power-law emission do not play a major role, since we focus on the brightest core of clusters where the ICM emission dominates in the RGS band. As shown in Fig.~\ref{fig:a85}, the typical instrumental background component at the target energy is just a few percent of the cluster emission. The background spectrum is nearly featureless in the plot. Such continuum components are hence unlikely to create any significant sharp feature at the target wavelength of 14.82~{\AA}. 

Another possible uncertainty in the continuum modelling is the column density ($N_{\rm H}$) of the absorbing Galactic ISM. The current $N_{\rm H}$ values are fixed to the total Galactic hydrogen column (\ion{H}{I} + H$_{\rm 2}$) published in \citet{willingale2013}. However, it is reported that the $N_{\rm H}$ measured with the X-ray data of galaxy clusters sometimes deviate from the Galactic hydrogen column. This might be partly due to the calibration uncertainties in the X-ray instruments and errors in the Solar abundance table \citep{schellenberger2015, deplaa2017}. How does a potential $N_{\rm H}$ error affect the spectral fitting at the target wavelength 14.82~{\AA}?

To test the effect of $N_{\rm H}$ uncertainty, we perform a new two-temperature fit to each object by allowing its $N_{\rm H}$ to vary freely. The Gaussian component at 14.82~{\AA} is excluded in the fits. The best-fit $N_{\rm H}$ values distribute around the Galactic values with a standard deviation of about 10$^{20}$ cm$^{2}$. We then shift the best-fit residuals to the source frame, and stack them in the same way as in Sect.~\ref{sect:results}. In Fig.~\ref{fig:inst}, we show that, for a narrow band 14.4~{\AA} $-$ 15.4~{\AA}, the stacked residual with the free $N_{\rm H}$ agree well with the original one with fixed $N_{\rm H}$, and the possible CX feature remains intact with the varying absorption.

Since RGS is a non-slit spectrometer, the spectral lines are broadened by the spatial extent of the source. For thermal plasma, it is determined by the projected emission measure distribution of the respective ion. Since the distributions of ions in a cluster might sometimes be different \citep{deplaa2006}, the resulting line widths thus
differ for different elements. However, the current standard spectral analysis tool cannot fit each element independently, it gives only an average line width (Sect.~\ref{sect:data}). The potential deviation between the average and the true width of strong emission lines might induce residuals mostly at the wing of the lines. Then question is whether or not this might explain the observed putative feature at 14.82~{\AA}.

To address the effect of line broadening, we force the line width to be a factor of two smaller/larger than the original values, and rerun the global fits for each object. The line width change is achieved by fixing the scale parameter \textit{s} to be twice/half of the best-fit values obtained in the original fits. In such a way, the broadening profile varies by a total factor of four, which is roughly consistent with the observed discrepancy between \ion{O}{VIII} and Fe lines reported in \citet{deplaa2006}. Following Sect.~\ref{sect:results}, we then blue-shift the new residuals obtained from the global fits to $z=0$, and stack them weighted by the counts. As shown in Fig.~\ref{fig:inst}, the two stacked residuals with scaled line broadening are in good agreement with each other at 14.4~{\AA} $-$ 15.4~{\AA}, and the positive residuals peaked at 14.82~{\AA} can be seen at the same level as the original fits. This means that the line broadening has little effect on the fits of the target band. This is probably owing to the lack of strong emission lines near 15~{\AA} for the selected clusters.       

\subsubsection{Bias due to atomic data errors}

Although the target \ion{O}{VIII} CX line is relatively well isolated in the spectra of hot clusters, it still
has some neighbouring weak lines. Could the putative feature at 14.82~{\AA} come from atomic data errors in the normal thermal emission? Here we investigate the atomic uncertainties
of the adjacent thermal lines. For the selected high-temperature clusters in our sample, there are only weak thermal lines, mainly from \ion{Fe}{XX} and \ion{O}{VIII}, in the proximity of the target CX line.

\ion{Fe}{XX} might emit two satellite lines at 14.824~{\AA} and 14.849~{\AA}, from the transitions of 2s2p$^{4}$ $^{4}$P$_{5/2}$ 
 - 2s$^2$2p$^2$($^3$P)3p $^4$D$_{7/2}$ and 2s2p$^{4}$ $^{4}$P$_{5/2}$ - 2s$^2$2p$^2$($^3$P)3p $^2$D$_{3/2}$, respectively. To address the atomic uncertainties, we compare the line emissivities in SPEX v3.03 and APEC v3.0.8. Fig.~\ref{fig:fexx} plots the SPEX and APEC 
 fluxes of these two satellite lines, scaled to the main \ion{Fe}{XX} line at 14.283~{\AA} from 2s2p$^{4}$ $^{4}$P$_{5/2}$ - 2s2p$^{3}$($^{5}$S)3s $^4$S$_{3/2}$ transition, as a function of the balance temperature. The relative differences between SPEX and APEC, which can be approximated as the atomic uncertainties, are $<$ 65\% for the 14.824~{\AA} line, and $<50$\% for the 14.849~{\AA} line. In the same figure, we show that line fluxes in both SPEX and APEC are much lower than 1\% of the local continuum. Though uncertain, these two satellite lines are apparently too weak to give any significant feature on the RGS spectrum. 
 
 The atomic error associated with the \ion{O}{VIII} Ly$\delta$ line emitted from thermal plasma also contributes to the systematic uncertainties. Here we compare the calculations by the CIE model in SPEX v3.03 and APEC v3.0.8. Fig.~\ref{fig:fexx} shows the total \ion{O}{VIII} Ly$\delta$ doublet fluxes normalized to the Ly$\alpha$ lines as a function of the balance temperature. The relative differences between SPEX and APEC are $<$ 4\% at 0.1 keV and 5\% at 2 keV. To create a line feature of 1\% of the local continuum, the \ion{O}{VIII} Ly$\delta$ flux must be underestimated by a factor of $>$ 2, which is much larger than the difference between SPEX and APEC. Hence, the thermal \ion{O}{VIII} lines are rather well calculated, and the induced systematic error to the target energy band is quite small.

\begin{figure}[!htbp]
\resizebox{0.9\hsize}{!}{\includegraphics[angle=0]{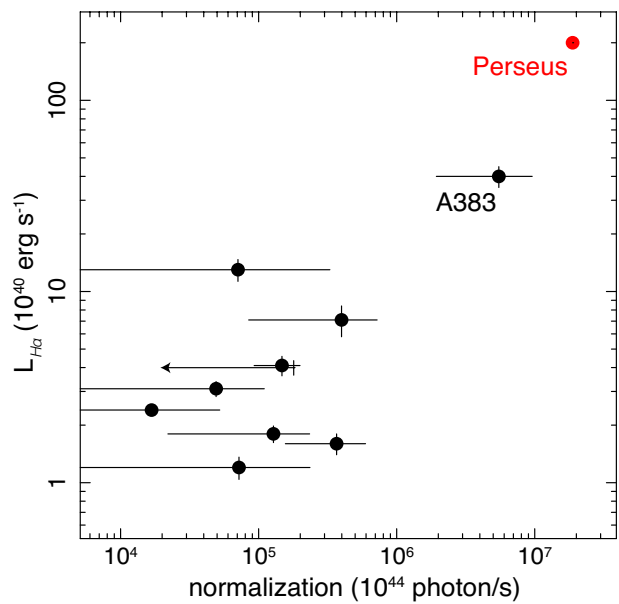}}
\caption{Measured normalizations of the Gaussian line at 14.82~{\AA} plotted against the H$\alpha$ line luminosities taken
from \citet{hamer2016}. As a reference, we also plot the expected \ion{O}{VIII} Ly$\delta$ line strength for the Perseus cluster, estimated based on the CX model 
that best-fits the \textit{Hitomi} data \citep{hitomi-atomic}. A solar abundance ratio is assumed since the Hitomi spectrum does not cover the oxygen band. 
The H$\alpha$ luminosity of the Perseus cluster is taken from \citet{conselice2001}. }
\label{fig:halpha}
\end{figure}

\begin{figure*}[!htbp]
\resizebox{0.6\hsize}{!}{\includegraphics[angle=0]{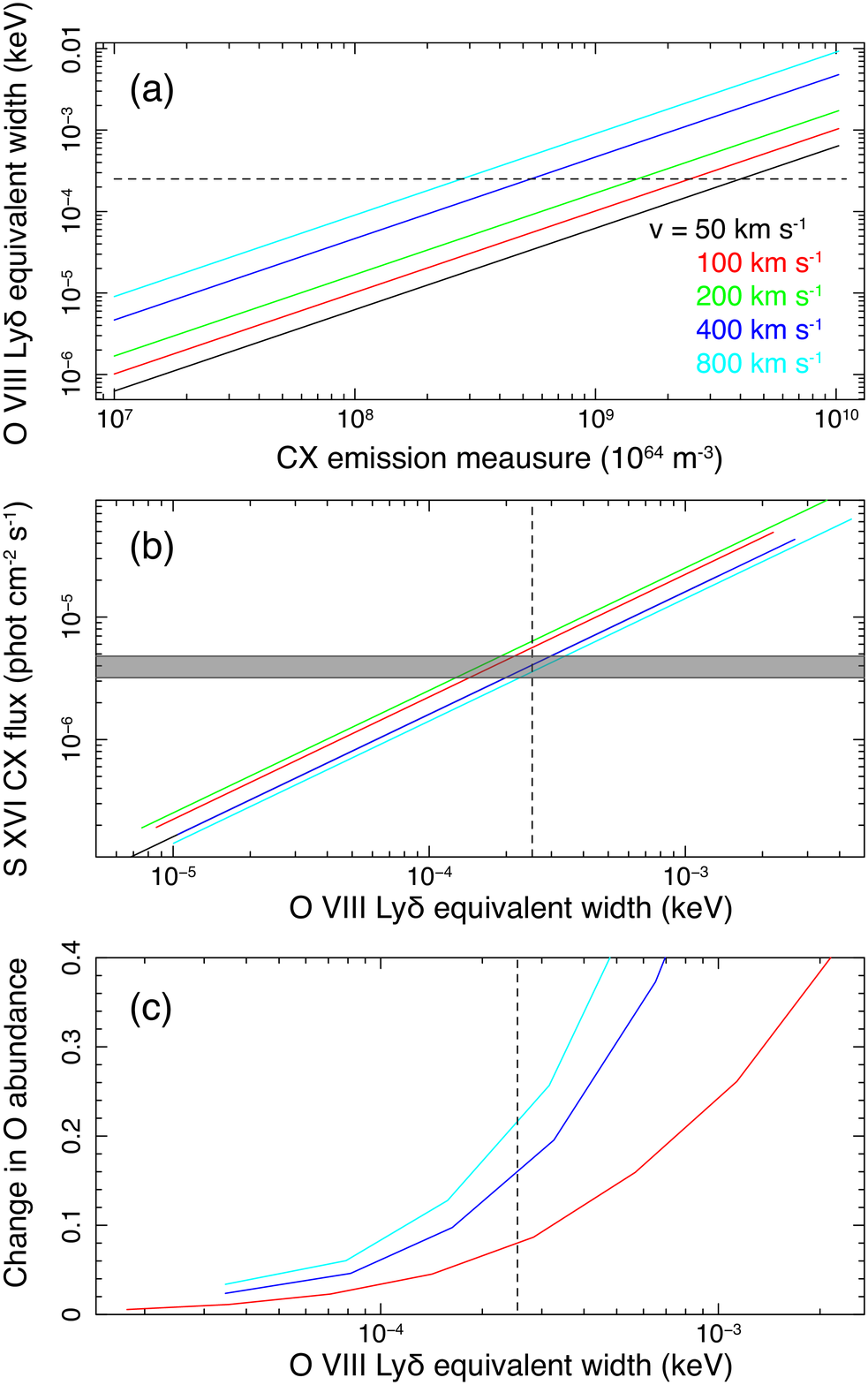}}
\caption{(a) Equivalent width of the \ion{O}{VIII} Ly$\delta$ line as a function of the CX emission measure for different collision velocities, calculated
based on the \textit{cx} model in SPEX. The background thermal component is taken from the best-fit two-temperature model for Abell~85. 
The observed sample-averaged equivalent width is shown by a dashed line in all panels. (b) Model calculation of \ion{S}{XVI} CX line
flux at $\sim 3.5$ keV as a function of \ion{O}{VIII} Ly$\delta$ equivalent width. The shadow region shows the observed flux
of the unidentified 3.5 keV line in \citet{bulbul2014}. (c) Fractional change in oxygen abundance as a function of \ion{O}{VIII} Ly$\delta$ equivalent width. The color schemes of panel (b) and (c) are the same as panel (a). }
\label{fig:disc}
\end{figure*}

\section{Discussion \label{sect:discussion}}

Using the high resolution spectroscopic data for a sample of massive galaxy clusters, we report a marginal detection of charge exchange feature at 14.82~{\AA} created by fully ionized oxygen interacting with neutral 
matter. Although the feature is weak, it cannot be easily explained 
either by the systematic uncertainties due to continuum modelling and instrumental broadening, or by the atomic uncertainties
of the adjacent thermal emission lines. The new CX feature poses a series of questions to be answered; (i) what is the origin 
of the possible CX emission? (ii) is the new \ion{O}{VIII} line consistent with the CX scenario proposed for the possible 3.5 keV line? and (iii) how does it potentially affect the previous measurements of the ICM abundance based on pure thermal modelling?

\subsection{Origin of the charge exchange emission}

Given the observed line profile and the derived parameters, it is likely that the CX emission originates from the cold gas clouds embedded in the hot ICM. The foreground solar wind charge exchange in the geocorona/heliosphere is negligible, since it can hardly form a line in the RGS spectrum due to the instrumental broadening. 

The cold gas clouds are often observed near the central galaxies in clusters \citep{conselice2001}, and/or in the wake of the member galaxies during their infall \citep{gu2013, gu2013b, yagi2015, gu2016b}. Since these neutral structures are completely immersed in the pool of highly-ionized 
plasma, the CX process is thus naturally expected at the interface, strongly affecting the ionization state of the interface.
To investigate the possible relation between the CX and the cold clouds, we compare in Fig.~\ref{fig:halpha} the best-fit normalization of the Gaussian component at 14.82~{\AA} for each cluster, with its H$\alpha$ luminosity measured with the VLT spectroscopic data \citep{hamer2016}. The H$\alpha$ data is chosen, since it illustrates the excitation (or recombination) process of the cold clouds, which might occur at a similar location as the CX. The plot shows that we can hardly conclude any significant relation between the possible CX and H$\alpha$ emission based on the current sample and data quality; a further study with deeper X-ray data is needed. One potential bias on the plot is that the H$\alpha$ data include only the central galaxies, while the RGS data might be also affected by the CX associated with the member galaxies in its field-of-view.

\subsection{Charge exchange model}

Here we provide a physical model for the possible \ion{O}{VIII} CX line. The CX flux is given by
\begin{equation}
F = \frac{1}{4 \pi D_{\rm l}^{2}} \int n_{\rm I} n_{\rm N} v \sigma_{\rm I, N}(v,n,l,S) dV,
\end{equation}      
\noindent where $D_{\rm l}$ is the luminosity distance of the object, $n_{\rm I}$ and $n_{\rm N}$
are the densities of ionized and neutral media, respectively, $v$ is the collision
velocity, $V$ is the interaction volume, and $\sigma_{\rm I, N}$ is the charge exchange cross section as a function of the velocity $v$ and the energy of the capture state characterized primarily by the quantum numbers $n$, $l$, and $S$. The cross section calculation is described in \citet{gu2016}. By defining emission measure as $\int n_{\rm I} n_{\rm N} dV$, we show in Fig.~\ref{fig:disc} the model calculation of equivalent width of \ion{O}{VIII} Ly$\delta$ line as a function of emission measure, for a collision velocity $v$ varying in the range of $50-800$ km s$^{-1}$. The thermal component used in the calculation is taken from the best-fit two-temperature model of Abell~85 (Fig.~\ref{fig:a85}). The ionization temperature and abundances of the CX component are assumed to be the same as the thermal plasma. To give the sample-average equivalent width of 2.5 $\times 10^{-4}$ keV at 14.82~{\AA}, the CX component would have an emission measure ranging from $4 \times 10^{67}$ cm$^{-3}$ for $v = 50$ km s$^{-1}$, to $3 \times10^{66}$ cm$^{-3}$ for $v=800$ km s$^{-1}$. Assuming an ICM density of $\sim 5 \times 10^{-2}$ cm$^{-3}$ at cluster cores \citep{zhuravleva2014} and a neutral gas density of $\sim 10$ cm$^{-3}$ in the molecular clouds \citep{heiner2008}, the effective interaction volume is then $\sim 5 - 15$ kpc$^{3}$. 

The estimated CX emission measure is generally consistent with the predicted value in \citet{gu2015}. The theoretical model in \citet{gu2015} is proposed to explain a possible weak emission line at $\sim 3.5$ keV from galaxy clusters reported in \citet{bulbul2014} and \citet{boyarsky2014}, by a CX reaction between bare sulfur ions and neutral atoms. It is hence naturally expected that the detections at the 3.5~keV and 14.82~{\AA} would boil down to the same CX source. To prove this, we plot in Fig.~\ref{fig:disc} the model calculation of the \ion{S}{XVI} flux at $\sim 3.5$ keV as a function of the \ion{O}{VIII} Ly$\delta$ equivalent width for different $v$. The sulfur ions are assumed to have the same ionization temperature and abundance as the oxygen ions. For the same \ion{O}{VIII} equivalent width,
the 3.5 keV flux increases with the collision velocity for $v < 300$ km s$^{-1}$, and decrease with velocity for larger $v$. This is because the CX capture spreads into more adjacent atomic levels of the sulfur ions at high velocity, and hence the line at 3.5 keV smears out. For the model yielding the \ion{O}{VIII} equivalent width of $2.5 \times 10^{-4}$ keV, the predicted \ion{S}{XVI} line flux at $\sim 3.5$ keV is $3.5 - 6.5 \times 10^{-6}$ photons cm$^{-2}$ s$^{-1}$, which agrees well with the observed value reported in \citet{bulbul2014}. This clearly shows that the possible \ion{O}{VIII} CX line is well in line with the model in \citet{gu2015} for the possible $\sim 3.5$ keV line. 

Here we compare our results with a previous work by \citet{walker2015}. By analyzing the stacked \textit{Chandra} CCD data of the X-ray/H$\alpha$
filaments in the Perseus cluster, \citet{walker2015} found that the X-ray spectra can be well fit with a CX component for the emission from the filament surface, 
together with a thermal component from the surrounding ICM. They reported a CX flux of $\sim 1.3 \times 10^{-13}$ ergs cm$^{-2}$ s$^{-1}$ in the $0.5-1.0$~keV band, 57\% of the 
total X-ray flux from the filament regions. This value is lower, by a factor of 5, than our estimate of the average CX flux in the same band, based on Eq.1 and the observed equivalent width 
of \ion{O}{VIII} Ly$\delta$ line. The difference might be explained by at least two facts: the selection of X-ray/H$\alpha$ filaments in \citet{walker2015} is far from complete (see their Figure 1), and the CX model used in their work \citep{smith2012} clearly differs from the one in our study \citep{gu2016}.

\citet{walker2015} also presented an \textit{XMM-Newton} RGS spectrum of the Centaurus cluster, showing a lack of significant oxygen lines predicted by their CX model. This is not surprising since
the CX features, if exist, are indeed rather weak in all objects of our sample (Fig.~\ref{fig:ew}), it must be difficult to report a significant detection based on the data from any individual cluster. 

\subsection{Effect on ICM abundance measurement}

\begin{figure}[!htbp]
\resizebox{\hsize}{!}{\includegraphics[angle=0]{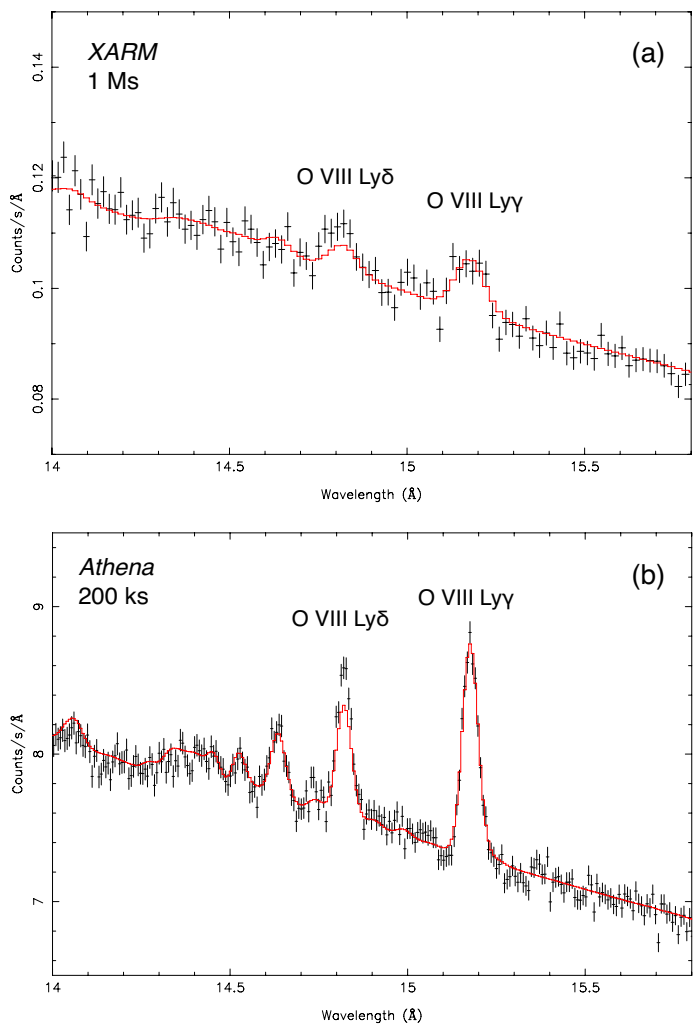}}
\caption{Simulated 1~Ms \textit{XARM} and 200~ks \textit{Athena} spectra of CIE + CX emission 
from an Abell~85-type object, fitted by a CIE model alone. } 
\label{fig:xarm}
\end{figure}

For such a newly-proposed component with a relatively weak spectral feature, the CX emission is often ignored in the ICM analysis. If the CX component does exist, 
it should have introduced a bias on the abundance measurement, as it produces strong Ly$\alpha$ lines which blends with the
thermal lines. By analyzing the \textit{Hitomi}
data of the Perseus cluster, \citet{hitomi-atomic} suggested that the Fe abundance is overestimated by $\sim 5$\% 
if the CX component is ignored in the fits. To estimate the possible impact on the oxygen abundance, we simulate the CX + CIE
spectrum in SPEX, and fit it with a pure CIE model. The input CIE model is taken from the best-fit model of Abell~85, 
multiplied by the observed line broadening profile. The input CX model has the same ionization balance and abundances as 
the CIE component. As shown in Fig.~\ref{fig:disc}, for the input CX \ion{O}{VIII} equivalent width of $2.5 \times 10^{-4}$ keV, the best-fit CIE abundance becomes higher than the input value by $\sim 8 - 22$\% for $v = 100-800$ km s$^{-1}$. It shows that
the model predicts a higher Ly$\alpha$/Ly$\delta$ ratio for a larger collision velocity. As reported in \citet{deplaa2017}, the measurement 
uncertainty on the oxygen abundance by the same RGS data of Abell~85 is 12\%. This
indicates that the bias on the ICM abundance caused by the possible CX component is marginally significant for the current instruments. 

This simulation is based on an object with a temperature of $3-4$ keV, and shows the typical level of systematic difference in abundance to expect. In real clusters with different temperatures, or if the spatial distribution of the CX component becomes very different from the thermal ICM, the effects on the abundance measurement will be different. The impact on other elements is expected to be comparable or relatively smaller.

The CX lines are calculated using the theoretical cross sections reported in \citet{janev1993}. As shown in \citet{gu2017},
the different theoretical approaches might sometimes differ quite a lot from each other, yielding a large systematic error on the line ratios. The laboratory data needed to verify the atomic codes are by far not complete. When instead using the results from multi-channel Landau-Zener method in \citet{mullen2017}, the \ion{O}{VIII} Ly$\alpha$/Ly$\delta$ ratio decreases by 70\% at $v=100$ km s$^{-1}$, which would hence lead to a further larger systematic difference on the oxygen abundance.

The possible charge exchange emission from diffuse astrophysical objects will be much better measured with future calorimeter 
missions. In Fig.~\ref{fig:xarm}, we show the simulated \textit{XARM} and \textit{Athena} spectra at the 
\ion{O}{VIII} Ly$\delta$ band. The input model consisting of the best-fit thermal model for Abell~85, and 
a CX component giving an equivalent width of $2.5 \times 10^{-4}$ keV at 14.82~{\AA}, is convolved with the 
\textit{XARM} and \textit{Athena} responses and fitted with a pure thermal model. The excess at 14.82~{\AA}
is detected at 5$\sigma$ for an exposure of 200~ks with the \textit{Athena} X-IFU. This indicates that to
characterize the charge exchange process in the intracluster space, we will need high spectral resolution spectra
from telescopes with a large effective area. For the \textit{XARM} spectrum, detecting the CX using the
\ion{O}{VIII} line alone becomes much more difficult; clearly we also have to investigate the CX features at
other energies, such as the \ion{S}{XVI} at $\sim 3.5$ keV and the \ion{Fe}{XXV} at $\sim 8.8$ keV, as those reported in \citet{hitomi-atomic} and \citet{hitomi3_5}. A more systematic simulation on the CX astrophysics with future X-ray missions will be present in an upcoming paper.

\section{Conclusion \label{sect:conclusion}}

We perform a systematic search for a charge exchange emission line at the rest-frame wavelength of 14.82~{\AA} in a RGS sample of 21 galaxy clusters.
The line is a characteristic feature that indicates strong physical interaction between bare oxygen ions and neutral particles. By fitting the thermal component and stacking the 
residuals, we do find a hint for a line-like feature at the target wavelength. The possible feature has a 2.8$\sigma$ significance above the local thermal component, corresponding
to an average equivalent width of 2.5$\times 10^{-4}$ keV. Although it only constitutes a marginal detection, it cannot be easily accounted for by either instrumental effects 
or atomic uncertainties.
If it is indeed a CX line from galaxy clusters, it would indicate exactly the same physical model as the one in \citet{gu2015}, which was proposed to explain an unidentified line
found earlier at $\sim 3.5$ keV. The model is also well in line with the possible CX component marginally detected with the \textit{Hitomi} spectrum of the Perseus cluster. Although the 
CX emission is expected to be weak, it implies a potential overestimation of the oxygen abundance by $8-22$\% in previous abundance measurements. A confirmation of this feature has to wait for spectroscopic observations with a future mission of higher sensitivity.

\begin{acknowledgements}

SRON is supported financially by NWO, the Netherlands Organization for
Scientific Research. 

\end{acknowledgements}

\bibliographystyle{aa}
\bibliography{main}
\end{document}